\RequirePackage{fix-cm}
\documentclass[smallextended]{article}       
 \usepackage{makeidx}
\makeindex

\usepackage{graphicx}
\usepackage{mathrsfs}
\usepackage{enumerate}
\usepackage{lscape}
\usepackage[utf8]{inputenc}
\usepackage[T1]{fontenc}
\usepackage{graphicx}
\usepackage{wrapfig}
\usepackage{graphicx}
\usepackage{amsfonts}
\usepackage{verbatim} 
\usepackage{amsmath}
\begin{document}
\title{The Vanishing Twist in the Restricted Three Body Problem
}



\author{Holger R. Dullin \\ School of Mathematics and Statistics, University of Sydney NSW 2006 \\
              \texttt{Holger.Dullin@sydney.edu.au}  \\         
       Joachim I. Worthington \\ School of Mathematics and Statistics, University of Sydney NSW 2006 \\
		\texttt{J.Worthington@maths.usyd.edu.au}
}


\renewcommand{\topfraction}{.85}
\renewcommand{\bottomfraction}{.7}
\renewcommand{\textfraction}{.15}
\renewcommand{\floatpagefraction}{.66}
\renewcommand{\dbltopfraction}{.66}
\renewcommand{\dblfloatpagefraction}{.66}
\setcounter{topnumber}{9}
\setcounter{bottomnumber}{9}
\setcounter{totalnumber}{20}
\setcounter{dbltopnumber}{9}


\maketitle

\begin{abstract}

This paper demonstrates  the existence of twistless tori and the associated 
reconnection bifurcations and meandering curves
in the planar circular restricted three-body problem. Near the Lagrangian equilibrium $\mathcal{L}_4$ 
a twistless torus is created near the tripling bifurcation of the short period family.
Decreasing the mass ratio leads to twistless bifurcations which are particularly prominent for rotation numbers $3/10$ and $2/7$.
This scenario is studied by numerically integrating the regularised Hamiltonian flow, and 
finding rotation numbers of invariant curves in a two-dimensional Poincar\'{e} map.
 
To corroborate the numerical results the Birkhoff normal form at $\mathcal{L}_4$ is  calculated to eighth order.
Truncating at this order gives an integrable system, and the rotation numbers obtained from 
the Birkhoff normal form agree well with the numerical results. A global overview 
for the mass ratio $\mu \in (\mu_4, \mu_3)$ is presented by 
showing lines of constant energy and constant rotation number in action space.


\end{abstract}

\maketitle
\section{Introduction}
\label{intro}

The circular restricted three body problem (CR3BP) describes the dynamics of a body of negligible mass (the \emph{test particle}) travelling in the gravitational field of two bodies with masses \(m_1\) and \(m_2\) (the \emph{primaries}).  The primaries are assumed to have circular orbits, and all three bodies are restricted to a single plane.  In the rotating frame of reference with the primaries fixed, there are five equilibria for the test particle, known as the \emph{Lagrange points}. In particular we are concerned with the triangular Lagrange equilibria \(\mathcal{L}_4\) and \(\mathcal{L}_5\) (which occur at the third point of equilateral triangles with the primaries). 
In celestial mechanics bodies whose orbit remains close to these Lagrange points are called Trojans, in particular in the Sun-Jupiter system.
It is well-known (see for example \cite{meyerhall}) that if \( \mu = \frac{m_2}{m_1+m_2}\leq \mu_1 = 0.5(1 -\sqrt{69}/9) \approx 0.0385\), then \(\mathcal{L}_4\) and \(\mathcal{L}_5\) are elliptic; for all parameter values considered in this paper, this condition is met. The monographs \cite{Bruno94} and \cite{Henon97} contain a wealth of information on the problem.
Here we are interested in invariant tori and their bifurcations near $\mathcal L_4$. 
The dynamics near $\mathcal L_4$ has also been studies extensively, see, e.g., \cite{DepHen70}, \cite{Schmidt74}, \cite{henrard02}, 
\cite{Richter01}.


At the triangular Lagrange equilibria for $\mu < \mu_1$ the linearised Hamiltonian has two pairs of pure imaginary 
eigenvalues $\pm i \omega_s$ and $\pm i \omega_l$ (corresponding to the short and long period families respective). 
The mass ratio $\mu$ for which $\omega_s/\omega_l = r > 1$ is
denoted by $\mu_r$, see, e.g., \cite{meyerhall}. 
According to the Lyapunov centre theorem, see, e.g., \cite{meyerhall}, the short period  family 
always exist for $\mu < \mu_1$ and has period $2\pi/\omega_s$ when approaching the origin.
When $r$ is not an integer then for $\mu_r$ the long period family exists and has period $2\pi/\omega_l$
when approaching the origin. For the range we are considering here $\mu \in (\mu_3, \mu_4)$ both the 
long and the short family exist. This range includes $\mu = 0.01215$ for the Earth-Moon system.

The computation of the Birkhoff normal form at  \(\mathcal{L}_4\) was first performed by hand by Deprit \& Deprit in 1967 \cite{deprit67}
up to order 4 (degree 2 in actions). 
They found the special value $\mu_c  \approx 0.0109$ for which the 
iso-energetic non-degeneracy condition of the KAM theorem does not hold at  \(\mathcal{L}_4\), 
and hence potentially stability could have been lost since the Hamiltonian is not definite at   \(\mathcal{L}_4\).
Meyer and Schmidt in 1986 \cite{meyer86} pushed the computation to order 6 using computer algebra.
With these results they established stability of  \(\mathcal{L}_4\) by KAM theory for all parameters in the elliptic range $\mu < \mu_1$,
except at lowest order resonances $\mu_2$, $\mu_3$ and $\mu_4$, but including $\mu_c$.

The vanishing of twist at the equilibrium point is not only a threat to KAM stability, but also signals the creation of a twistless torus nearby.
Standard iso-energetic KAM theory fails near a twistless torus. There are KAM theorems with weaker 
non-degeneracy assumptions that fix this problem \cite{Ruessmann87}. 
But in any case, the passage of the twistless torus through a rational rotation number under variation of a parameter (e.g.~the energy)
gives rise to  twistless bifurcations. Such bifurcations have first been studied in area preserving maps by Howard \cite{HowHoh84,HowHum95}, 
and a normal form was derived in \cite{Simo98}. Results on break-up of twistless curve and a review of the literature can be found in \cite{Wurm05}.
 
To analyse twistless bifurcations in an integrable or near-integrable map the most useful presentation of 
 the dynamics of the map is in the form of (approximate) action-angle variables $(I,\theta)$, so that the unperturbed dynamics
 is $I' = I$ and $\theta'  = \theta + W(I)$. At a twistless torus $I=I^*$ by definition the rotation number $W$ 
 has a critical point, $W'(I^*) = 0$. When the critical value $W(I^*)$ passes through a rational value
for a typically perturbed map a twistless bifurcation occurs.
 The paper \cite{dullin00} on generic twistless bifurcations demonstrates that
 for a family of maps with a fixed point with rotation numbers in $[1/6, 1/2]$ the 
 twist of the fixed point must vanish for some $W \not = 1/3$.
Vanishing twist at the origin corresponds to a critical point of $W$ at the origin, i.e. $W'(0) = 0$.
When the critical point of $W$ moves to positive $I$ under variation of a parameter a twistless curve is created.
Another source of a twistless curve may a saddle-centre bifurcation, 
as shown in \cite{DI03a}, where the universal features of this bifurcation were also studied.
For example, in the area preserving H\'enon map a twistless curve is created in a saddle-centre bifurcation of 
a period three orbit near a fixed point with rotation number 1/3 \cite{DI03a}.

Twistless bifurcations shown to exist for area preserving maps naturally also appear in 
Hamiltonian flows, simply because a Poincar\'e section of the flow at fixed energy gives a 2-dimensional
area preserving map. Thus we should expect to see twistless bifurcations in a Hamiltonian flow, 
since the energy $E$ may serve as the family parameter. In the CR3BP there is the additional external
parameter $\mu$, so that one expects to see twistless bifurcations along 1-dimensional lines in parameter 
space of $E$ and $\mu$.

The structure of the paper is as follows. The Hamiltonian and its regularisation is reviewed  in the next section.
In section~\ref{sec:Poincare} we present numerically computed Poincar\'e sections and the corresponding 
numerically computed $W$ curves. Our numerically computed results show that there are prominent
twistless bifurcations for $W = 3/10$ and $W = 2/7$ along one-parameter families in 
$(E, \mu)$ space. These occur in the island of stability in the Poincar\'e section that has the short 
period orbit at its centre. The line of twistless bifurcations in parameter space trace closely 
the lines where the short period orbit has rotation number $3/10$ and $2/7$, respectively.
Then we show that the numerical results can be analytically explained using a high order 
non-resonant Birkhoff normal form at the equilibrium \(\mathcal{L}_4\).
The presence of an equilibrium point is a feature of the flow that cannot be present 
in a family of maps. It serves as an anchor for the computation of an analytic approximation 
from which an approximation to $W(I)$ can be found. The normal form reveals the full complexity 
of the bifurcations of invariant tori around the short period family for 
$\mu \in (\mu_4, \mu_3)$ in the C3RBP.

%
%

\renewcommand{\arraystretch}{1.5}
\begin{table}
\centering
\begin{tabular}[c]{| c | c | c | c | c | c | c | c |}
	\hline
	\(\mathbf{r}\) & \({4}\) & \(\frac{11}{3}\) & \(\frac{7}{2}\)& \(c\)  & \(\frac{10}{3}\) & $\approx \frac{16}{5}$ & \({3}\) \\
	\hline
	\(\boldsymbol{\mu}_\mathbf{r}\) & 0.00827 &  0.00964 & 0.01045 & 0.01091 &  0.01135 & 0.01215 & 0.01351\\
	\hline
\end{tabular}
\caption{\(\mu_r\) for rational values of \(r\) between the \(1/4\) and \(1/3\) resonances, to five decimal places. 
\(\mu_c\) is the value where the 
twist vanishes at $\mathcal L_4$.
Rotation number $5/16$ approximately occurs for the Earth-Moon system.}
\label{tab:muvals}
\end{table}

\section{The Planar Circular Restricted Three Body Problem}


\label{sec:threebodyproblem}

We construct the Hamiltonian describing the dynamics of the test particle as in \cite{koon08}:
\begin{equation}
\label{eq:rotatinghamiltonian}
	\mathscr{H}(x,y,p_x,p_y)
		=\frac{1}{2}p_x^2+\frac{1}{2}p_y^2+p_xy-p_y\left (x+\mu-\frac{1}{2} \right ) - \frac{1-\mu}{r_1}-\frac{\mu}{r_2}+s,
\end{equation}
\begin{equation}
	r_1=\sqrt{(x+\frac{1}{2})^2+y^2}, \qquad r_2=\sqrt{(x-\frac{1}{2})^2+y^2}, \qquad s = \frac{3 + \mu(\mu-1)}{2} \,.
\end{equation}
The Hamiltonians are normalised by $s$ so that the energy of a stationary test particle at \(\mathcal{L}_4\) is \(E=0\). 
The Hamiltonian has been scaled so the total mass of the primaries is \(1\) unit, and the distance between the primaries is also \(1\) unit. We take a rotating frame of reference, with the primaries fixed; the primary of mass \(\mu\) at \((-1/2,0)\)  and  \(1-\mu\) at \((+1/2,0)\) for \(\mu\in(0,1/2]\), the \emph{mass ratio}. The other parameter of the system is \(E=\mathscr{H}\), the constant value of the Hamiltonian. 
Our $E$ is related to the usual \emph{Jacobi integral} \(C=-2(E-s)\).

We then introduce complex coordinates \(z=x+iy \in \mathbb{C}\), \(p_z=p_x-ip_y \in \mathbb{C}\). Now the transformed Hamiltonian is
\begin{equation}
	\label{eq:complexhamiltonian}
	\mathscr{H}(z,p_z)=\frac{1}{2}|p_z|^2+\operatorname{Im} \left ((z+\frac{1}{2}-\mu)p_z \right )
			-\frac{1-\mu}{|z+\frac{1}{2}|}-\frac{\mu}{|z-\frac{1}{2}|}+s.
\end{equation}

To regularise and remove the singularities at \(z=\pm 1/2\), we take the Thiele transformation as described in \cite{broucke63}, \cite{broucke65}, \cite{szebehely67}  
\begin{equation}
	\label{eq:thiele}
	z=\frac{1}{2}\cos{w},\;\;\;\;p_z=\frac{-2p_w}{\sin w}.
\end{equation}
The new Hamiltonian is
\begin{equation}
	\label{eq:regularised}
	\mathscr{H}(w,p_w)=2\left | \frac{p_w}{\sin w} \right | ^2 -\operatorname{Im}\left((\cos w +1-2\mu)\frac{p_w}{\sin w}\right)-\frac{2-2\mu}{|\cos w+1|}-\frac{2\mu}{|\cos w -1|}+s.
\end{equation}

Our final step is to take the Hamiltonian into extended phase space with time \(t\) and the fixed energy \(E=\mathscr{H}(w,p_w)\) now coordinates of our system. This construction is detailed e.g.~in \cite{szebehely67}. 
The time transformation \({\mathrm{d}t}/{\mathrm{d}\tau}=| \sin^2 w |/4\) (where \(\tau\) is our new ``time'' variable) then removes both singularities:
\begin{equation}
	\mathscr{K}(w,t,p_w,E)=\frac{1}{4}|\sin^2 w|(\mathscr{H}(w,p_w)-E).
\end{equation}
This Hamiltonian is identically \(0\) along test particle trajectories. Expanding gives the regular Hamiltonian used for the numerical integration
\begin{equation}
\begin{split}
	\mathscr{K}(w,t,p_w,E)	&=\frac{1}{2}|p_w|^2-\frac{1}{4}\operatorname{Im}((\cos w +1-2\mu)p_w\overline{\sin w})\\
					& -\frac{1-\mu}{2}|1-\cos w| - \frac{\mu}{2}|1+\cos w|-\frac{1}{4}|\sin^2 w|(E-s).
\end{split}
\end{equation}
This corresponds to the regularised Hamiltonian derived by Broucke in \cite{broucke65} in the case \(n=1\).

In real coordinates \(u,v,p_u,p_v\in \mathbb{R}\) (such that \(w=u+iv\), \(p_w=p_u-ip_v\)
so $u, p_u$ and $v, p_v$ are canonically conjugate pairs) the Hamiltonian is
\begin{align}
	\mathscr{K}(u,v,t,p_u,p_v,E)=&\frac{p_u^2}{2}+\frac{p_v^2}{2}
		+\frac{1}{2}((1 - 2\mu)\cos u - \cosh v)\\	
\nonumber		&+\frac{1}{4}\sin u ( (1- 2\mu) \cosh v+\cos u ) p_v\\
\nonumber		&+\frac{1}{4}\sinh v ((1  - 2\mu )\cos u+\cosh v) p_u\\			
\nonumber		&+\frac{1}{8}(\cos(2u)-\cosh(2v))(E-s).
\end{align}
Regularisation is not essential for the orbits we are mostly interested in, since they are confined to a neighbourhood 
of ${\mathcal L}_4$. However, for numerical purposes it is quite convenient to have regularised equations, so that orbits 
that have near collisions do not cause numerical problems or slow down the integration.

\section{Poincar\'e sections}
\label{sec:Poincare}

\begin{figure*}
\centering
\fbox{\includegraphics[ width=0.48\textwidth]{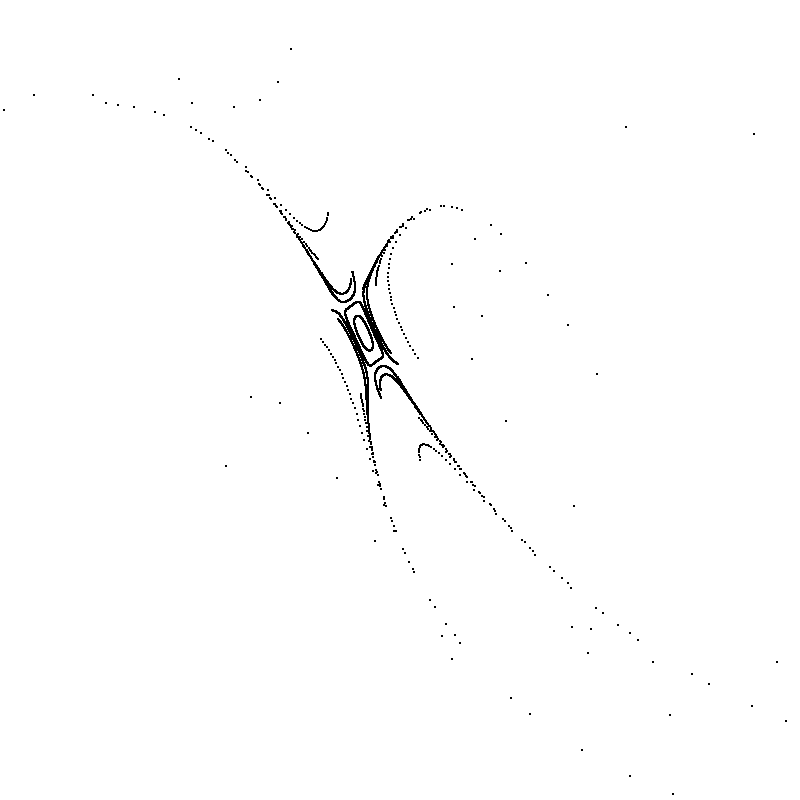}}
\fbox{\includegraphics[ width=0.48\textwidth]{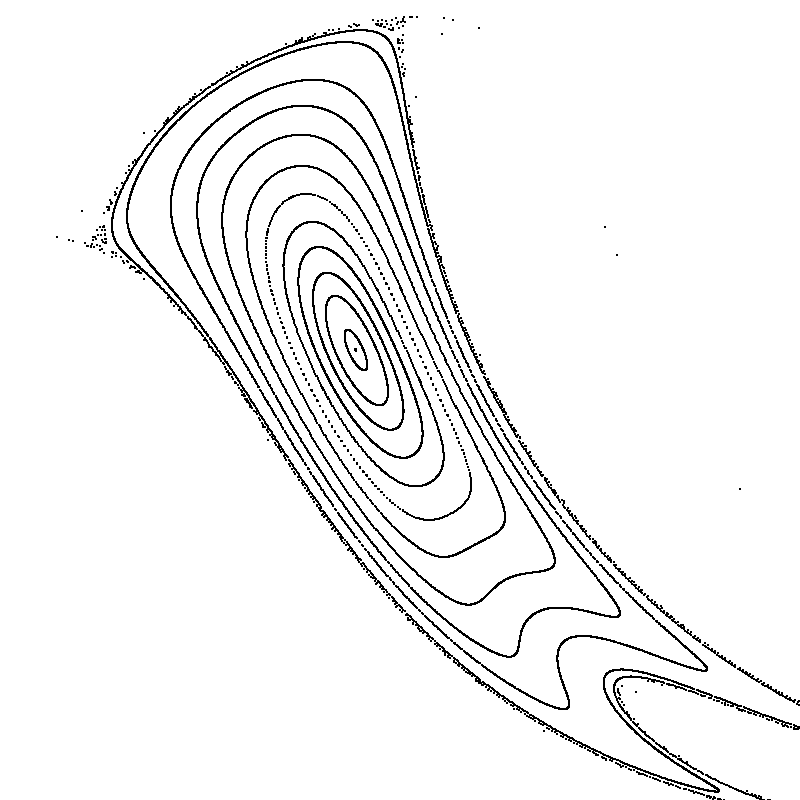}}
\fbox{\includegraphics[ width=0.48\textwidth]{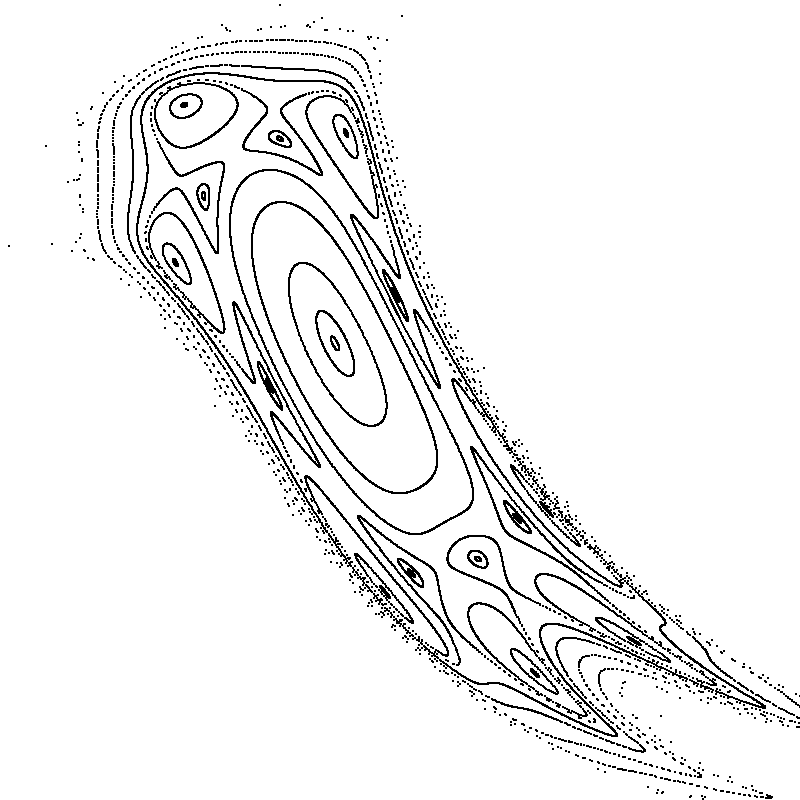}}
\fbox{\includegraphics[ width=0.48\textwidth]{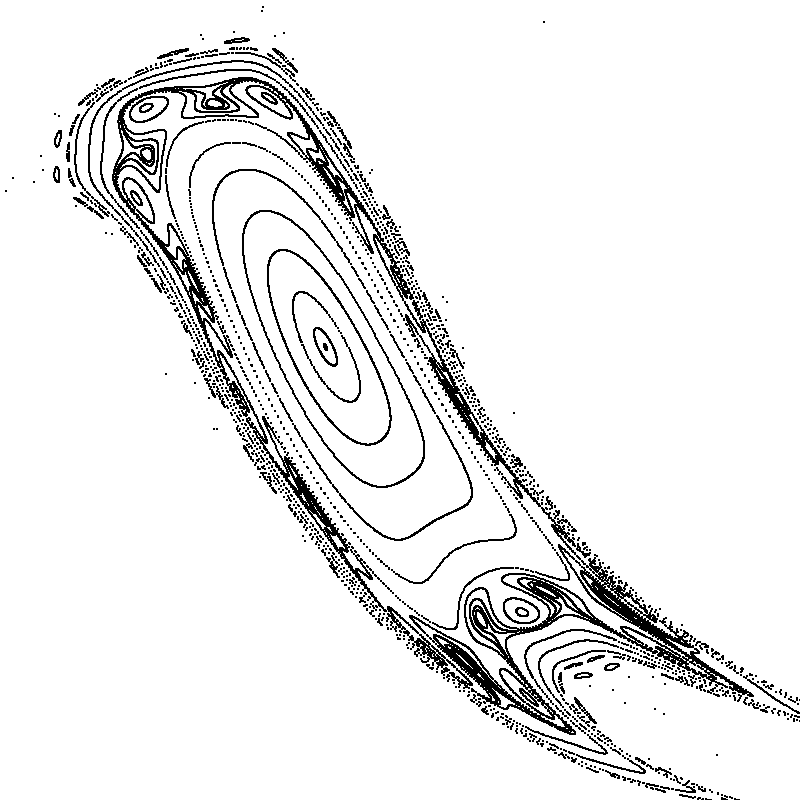}}
\fbox{\includegraphics[ width=0.48\textwidth]{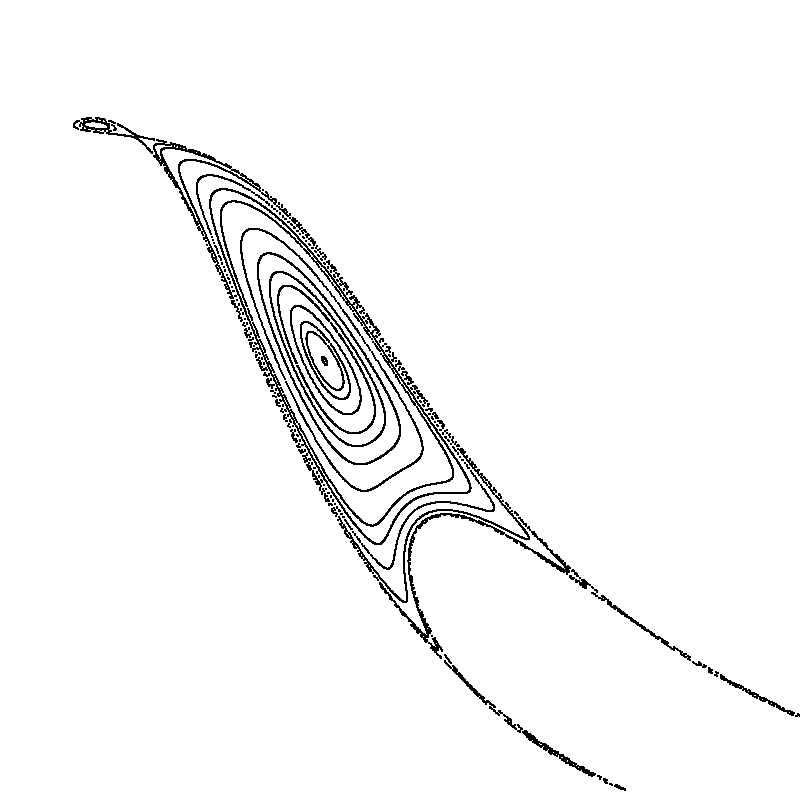}}
\fbox{\includegraphics[ width=0.48\textwidth]{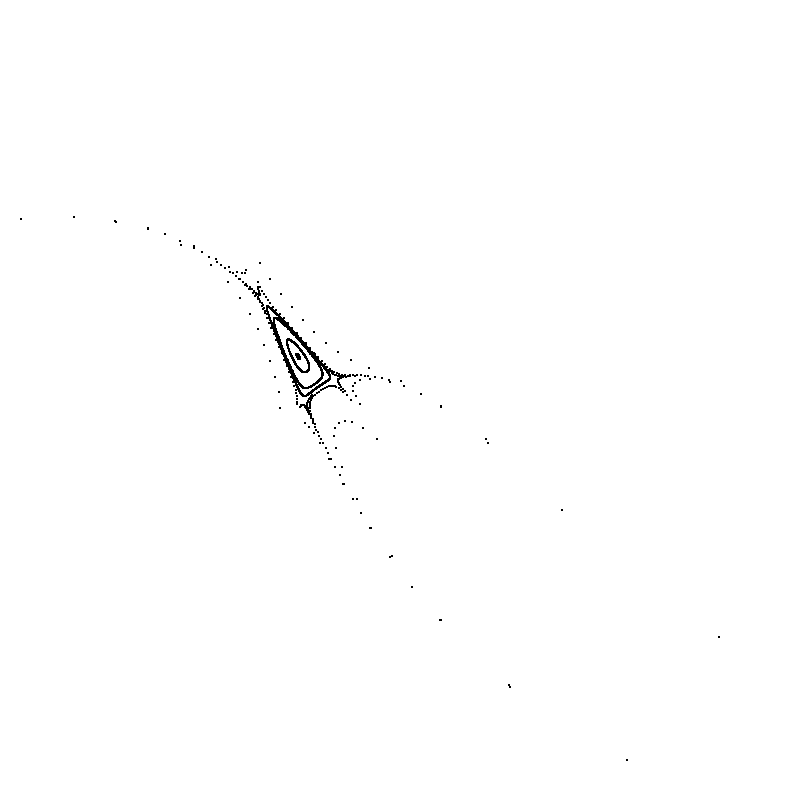}}
\caption{Poincar\'{e} Sections for \(E=0.02\) in \((a,p_a\) space. The area shown is \(a\in[0.72,0.89],\;p_a\in[-0.22,0.05]\). From left to right, top to bottom: \(\mu=0.007480\) (near \({1}/{4}\) resonance), \(\mu=0.0086\), \(\mu=0.009165\) (near \({2}/{7}\) reconnection bifurcation), \(\mu=0.009723\) (near \({3}/{10}\) reconnection bifurcation), \(\mu=0.0104\) (near period 3 saddle-centre bifurcation), 
\(\mu=0.011283\)  (near \({1}/{3}\) resonance). }
\label{fig:psections1}     
\end{figure*}

\begin{figure*}[t]
\centering
\fbox{\includegraphics[ width=0.48\textwidth]{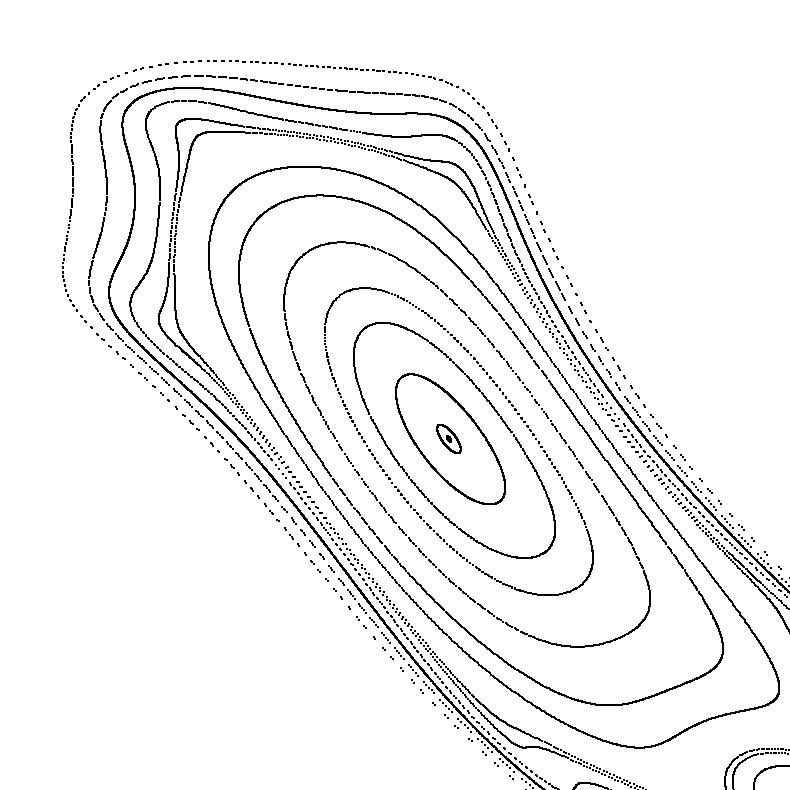}}
\fbox{\includegraphics[ width=0.48\textwidth]{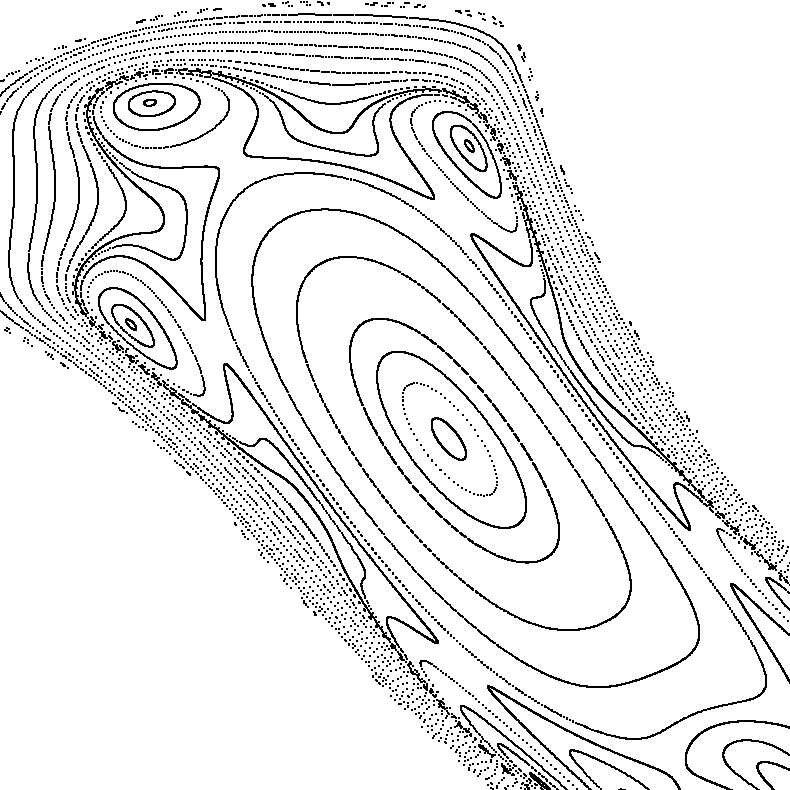}}
\fbox{\includegraphics[ width=0.48\textwidth]{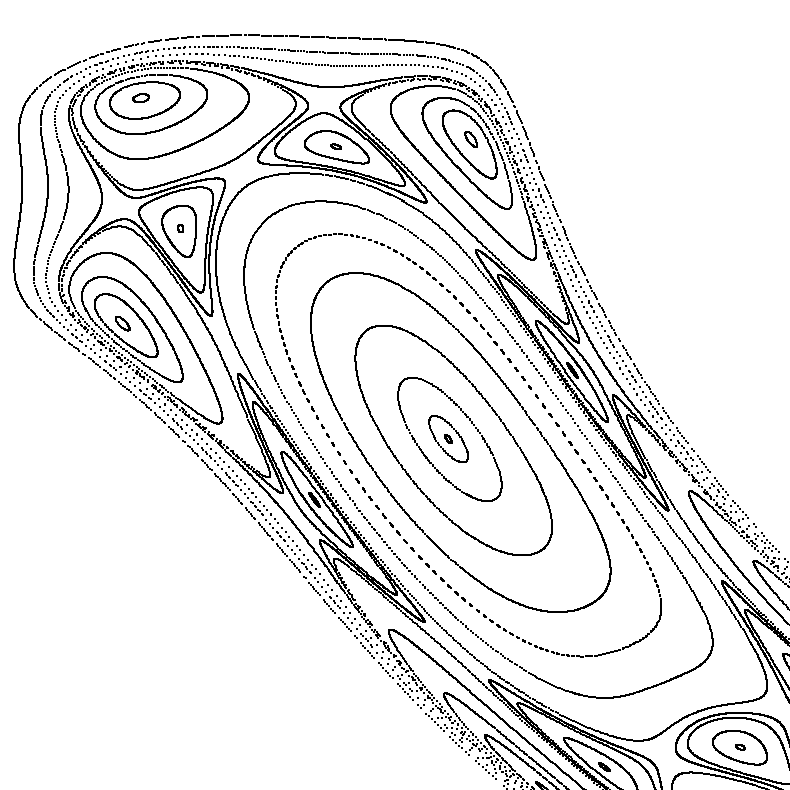}}
\fbox{\includegraphics[ width=0.48\textwidth]{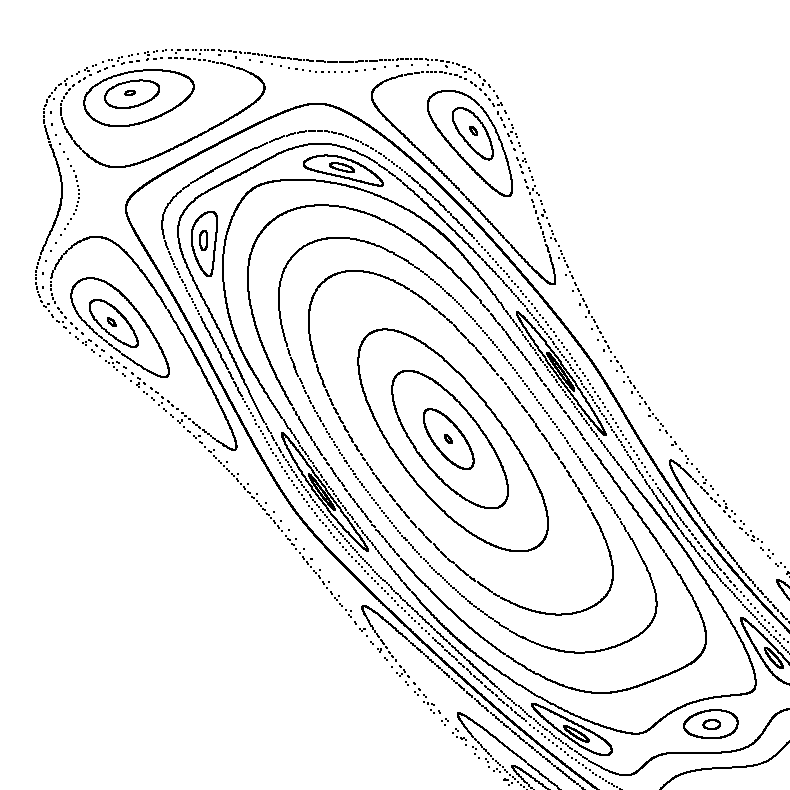}}
\caption{Details of the \({2}/{7}\) reconnection bifurcation shown in Figure \ref{fig:psections1} (\(E=0.02\)). This figure is also in \((a,p_a\) space, the area shown is \(a\in[0.74,0.83],\;p_a\in[-0.14,0.03]\). From left to right, top to bottom: \(\mu=0.009151\), \(\mu=0.009161\), \(\mu=0.009165\), \(\mu=0.009171\).}
\label{fig:twochains}     
\end{figure*}

The equations of motion derived from the Hamiltonian above were numerically integrated using a standard fourth order Runge-Kutta method. 
To visualise the dynamics we define a  \emph{Poincar\'{e} first return map}.
A general recipe to construct Poincar\'e sections that capture at least every bounded orbit 
was given in \cite{DW95}, and has been applied to the restricted problem in \cite{Richter01}.
Since here we are interested in orbits near $\mathcal L_4$ an know where they are located 
we use a simpler traditional approach, in which 
 the Poincar\'{e} section defined by the line in configuration space that passes through the heavier primary at \(z=-{1}/{2}\), and \(\mathcal{L}_4\). 
In \(z\)-coordinates, this section is a straight line; in \(w\)-coordinates, it is defined by the curve
\begin{equation}
	\sqrt{3}\operatorname{Re}(\cos w)-\operatorname{Im}(\cos w)+\sqrt{3}=0.
\end{equation}
We calculate the Poincar\'{e} map using the algorithm described in \cite{henon82}, with the modification described in \cite{palaniyandi09}. New coordinates defined by
 \begin{equation}
 	a=2\operatorname{Re}(z)+1,\;\;\;p_a=\frac{1}{2}\operatorname{Re}(p_z)-\frac{\sqrt{3}}{2}\operatorname{Im}(p_z)	
 \end{equation}
 are the plotted, as per \cite{sandor00}. 
 We create phase portraits such as Figures \ref{fig:psections1} and \ref{fig:twochains} of the two dimensional Poincar\'{e} section to identify the tori and their bifurcation visually. 
The program used for this is coded in C++ and OpenGL, and can quickly create images for large ranges of \(\mu\) and \(E\).
The computations are performed in the regularised coordinates, but the resulting Poincar\'e sections are transformed back 
into the cartesian variables \((a,p_a)\) defined above.  

For $E>0$ the short period family gives a fixed point in the Poincar\'e section. The island of stability surrounding this 
fixed point typically contains a large set of invariant curves, corresponding to tori in the full system.
See Figure \ref{fig:psections1} and Figure \ref{fig:twochains} for phase portraits of the Poincar\'{e} map, showing these invariant curves. 
The phase portraits in Figure \ref{fig:psections1} show the area \(a\in[0.72,0.89],\;p_a\in[-0.22,0.05]\).
The phase portraits in Figure \ref{fig:twochains} show details from Figure \ref{fig:psections1} in the area \(a\in[0.74,0.83],\;p_a\in[-0.14,0.03]\).
The main objective of this paper is to study the rotation number $W$ of this family of tori for $E>0$ close to 0
and $\mu \in (\mu_4, \mu_3)$, and identify reconnection bifurcations that correspond to a maximum of $W$
passing through a low order rational number.

\setlength{\fboxrule}{0.5pt}
\setlength{\fboxsep}{0.5pt}

A typical one-parameter family illustrating the main result is shown in Figure \ref{fig:psections1}. 
The energy parameter \(E\) is fixed, and Poincar\'{e} sections are created at various choices of the mass parameter \(\mu\). Much of the structure described  in \cite{dullin00} is clear; as \(\mu\) {\em decreases}, we observe the \({1}/{3}\) resonance, 
then the period 3 saddle-centre bifurcation,
then the structure associated with a \({3}/{10}\) reconnection bifurcation (the reconnection bifurcations are observed as two chains of ``islands'' colliding), 
then a similar twistless bifurcation with rotation number \({2}/{7}\) and then the island shrinks to the \({1}/{4}\) resonance. 
Since the twistless bifurcation occurs for very small interval in $\mu$ it is 
unlikely to observe these bifurcations for randomly chosen parameters. The sixth digit needs to be adjusted to locate the \(2/7\) 
reconnection, see Figure~\ref{fig:twochains}. Nevertheless, the corresponding structure in phase space 
for low order rationals is quite prominent.
The \(\mu=0.0086\) picture in   Figure \ref{fig:psections1} is very close to the ${3}/{11}$ twistless bifurcation, 
but not sufficiently close to show this structure.

Of particular note is the appearance of the \({2}/{7}\) reconnection bifurcation structure; for the H\'{e}non map example studied in \cite{dullin00}, the twistless curve collides with the fixed point between the parameter value where the rotation number is \({2}/{7}\) and at \({3}/{10}\). Indeed, the CR3BP also demonstrates a fairly prominent \({3}/{11}\) reconnection bifurcation. Higher order rational rotation numbers also create twistless bifurcations, but
they are not included here as they are hardly visible (in both phase space and parameter space) and are qualitatively much the same as the ones included. 
It seems to be the case that the twistless torus persists all the way to the \({1}/{4}\) resonance, but an analytical study of the corresponding 
resonant normal form is needed to establish this.

Figure \ref{fig:twochains} shows more detail of the \({2}/{7}\) reconnection bifurcation. Again, \(E=0.02\) is fixed while \(\mu\) is varied.

\begin{figure*}
\centering
 \fbox{\includegraphics[ width=0.44\textwidth]{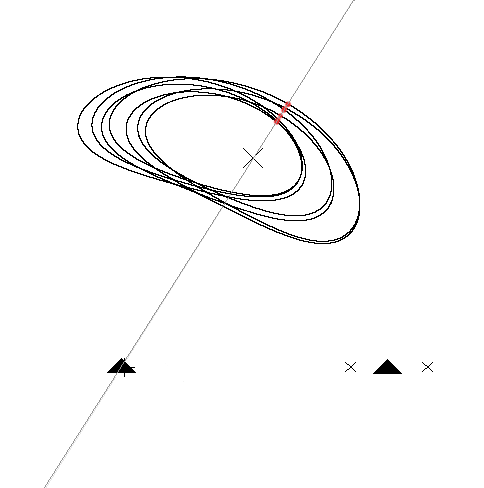}}
 \fbox{\includegraphics[ width=0.44\textwidth]{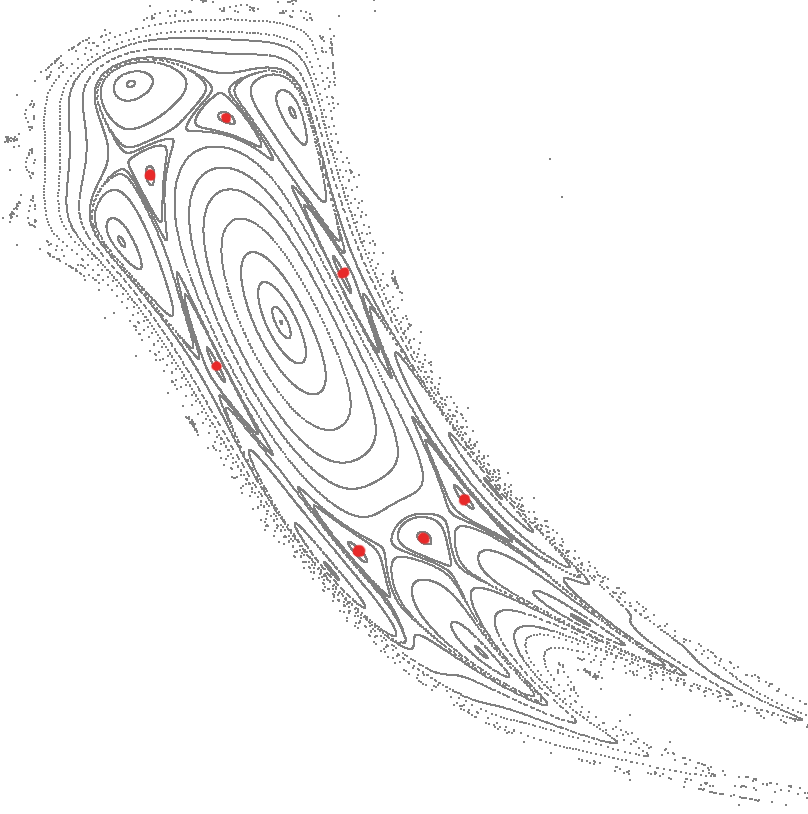}}
\caption{Left: A particular orbit for the parameter values \(E=0.02\), \(\mu=0.009165\) (shown in rotating coordinates). The triangles correspond to the primaries, the crosses correspond to the Lagrangian points, and the grey line is the Poincar\'{e} section. This orbit corresponds to the periodic orbit coloured red in the Poincar\'{e} section (right), and the parameters \(E\) and \(\mu\) are chosen such that the \({2}/{7}\) chains collide in a reconnection bifurcation.}
\label{fig:zspaceorbit}
\end{figure*}

Figure \ref{fig:zspaceorbit} shows a particular orbit which occurs at the \({2}/{7}\) reconnection bifurcation. The orbit shown corresponds to the mapping with period 7 indicated on the section at right. Although it is difficult to see, the orbit crosses the Poincar\'{e} section (indicated by a light grey line through the left primary and \(\mathscr{L}_4\)) 7 times in each direction, which accords with the map having period seven for this orbit.

\begin{figure*}
\centering
\includegraphics[width=13cm]{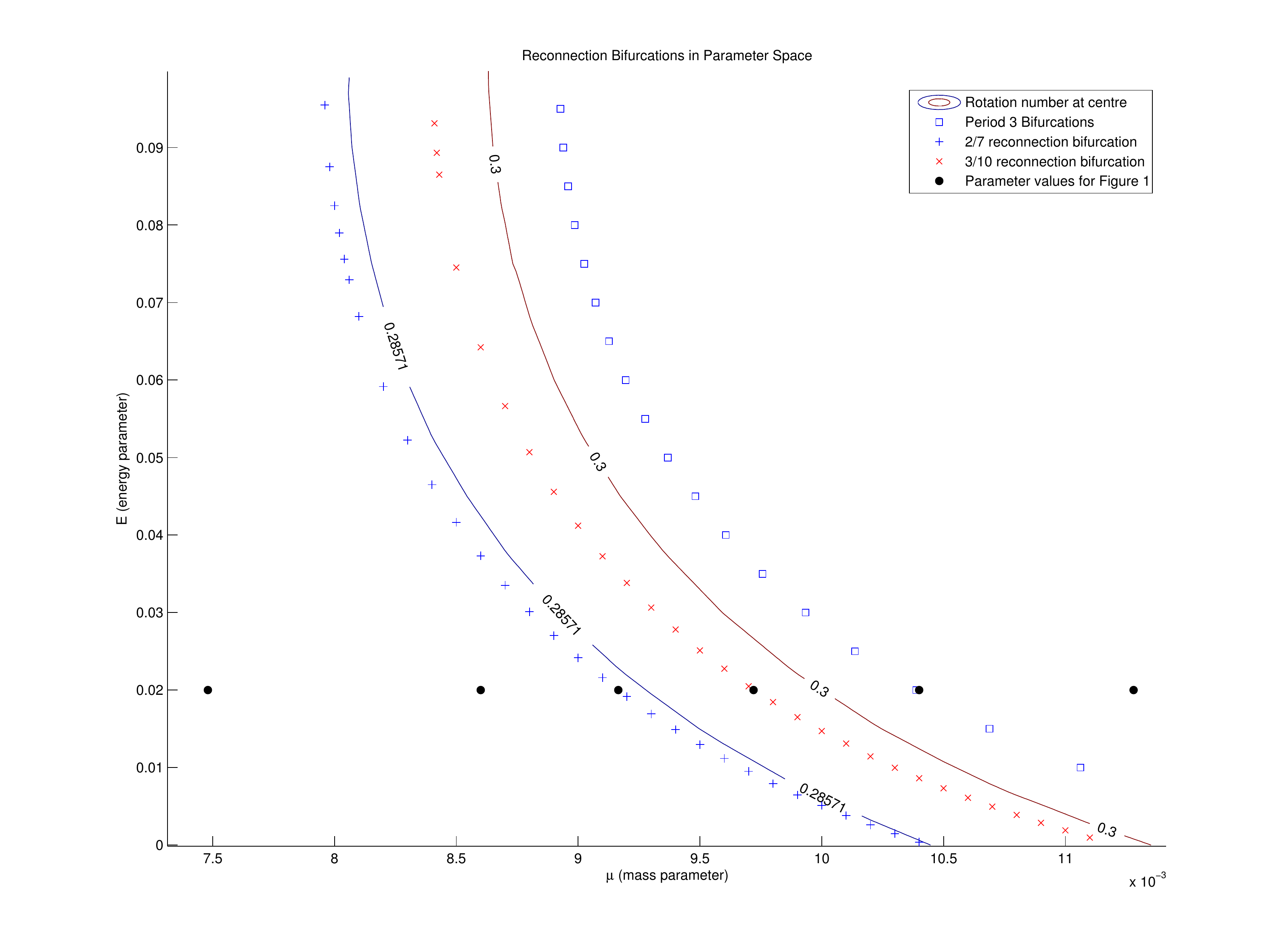}
\caption{Parameter values where \({2}/{7}\) and \({3}/{10}\) reconnection bifurcations occur. Parameter values where period 3 saddle-centre bifurcation occur are shown as squares. Also shown are contours for where the (approximate) rotation number at the fixed point is \({2}/{7}\), and \({3}/{10}\). Compare these contours with the analytical results in Figure \ref{fig:analyticalrotation}.
The black dots indicate the parameter values used in Figure~\ref{fig:psections1}.}
\label{fig:collisions}     
\end{figure*}

We have demonstrated that the system has twistless  bifurcations for a fixed \(E\) value as \(\mu\) is varied. The next step is to take a broader view of the parameter space. We plot the points where the \({2}/{7}\) and \({3}/{10}\) reconnection bifurcations occur in $(E, \mu)$ parameter space in Figure \ref{fig:collisions}. This figure also includes contours where the rotation number at the fixed point of the Poincar\'{e} map is \({2}/{7}\), and \({3}/{10}\) according to our numerical computations (cf. the analytical approximations in Figure \ref{fig:analyticalrotation}).
Recall that for $E >0$ the fixed point of the Poincar\'e map corresponds to the short period family.
Figure \ref{fig:collisions} shows that for the energy interval  \( E \in ( 0, 0.1] \) 
whenever the short period family has rotation number $3/10$ or $2/7$ then for slightly smaller $\mu$ and/or $E$ there is 
a twistless bifurcation. In particular every one-parameter family with fixed energy in this range exhibits these bifurcations.

\begin{figure*}
\centering
\includegraphics[width=0.84\textwidth]{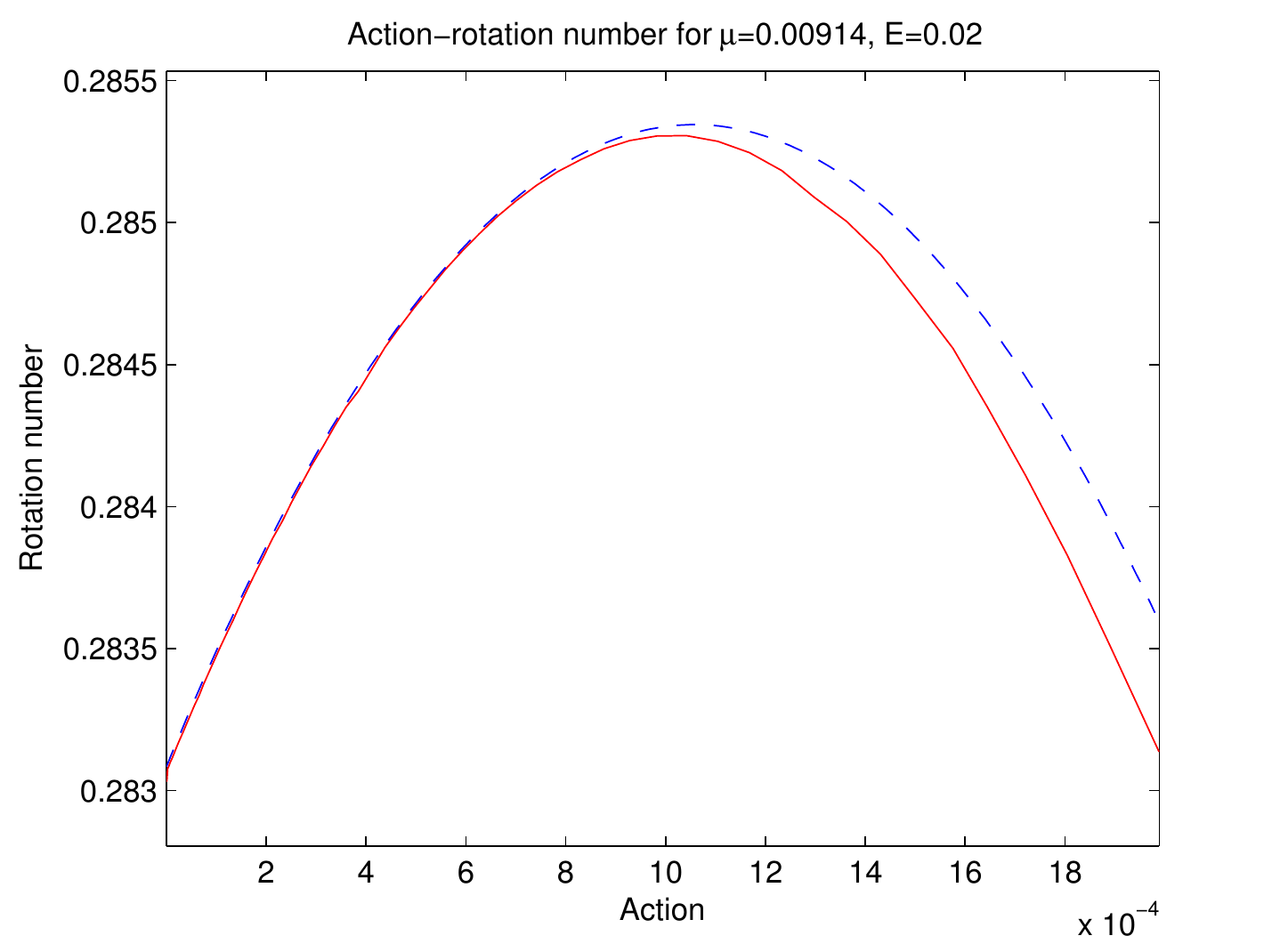}
\caption{Action vs rotation number for \(\mu=0.00914\), \(E=0.02\). The full red line shows the calculated rotation number, while the dashed blue line shows the analytical approximation. The curves agree for low values of the action, drifting further apart as the action increases. The critical value of this function corresponds to the twistless curve.}
\label{fig:actionrotation}
\end{figure*}

For fixed $E$ and $\mu$  invariant curves in the island can be parametrised by the 
area they enclose, which is given by the action $I_l$. 
Typical plots of $W(I)$ for fixed $E$ and $\mu$ 
are shown in Figure~\ref{fig:actionrotation}.  These figures are calculated using the algorithm described in \cite{efstathiou01}. Although these plots are not always accurate (particularly around the low order rationals), they provide a good guide to finding the bifurcations associated with the vanishing twist.
A twistless bifurcation occurs when the maximum of the rotation function 
$W(I^*)$ passes through a rational number.

\section{Vanishing Twist in flows}
\label{sec:vanishingtwist}

Here we briefly review vanishing twist in the context of a smooth integrable system with 2 degrees of freedom 
given in terms of a Hamiltonian $H(I_1, I_2)$; that is, a function of the actions only. For a review of the somewhat 
simpler situation of an integrable map see \cite{dullin00}.
First define the frequencies $\omega_i(I_1, I_2) = \partial H/\partial I_i = H_i$, and the frequency ratio 
$W(I_1, I_2) = H_1/H_2$. 
The 2-dimensional action space is the image of the 4-dimensional phase space under the action map. 
The preimage of a point in action space is 2-dimensional invariant torus in phase space. 
We think of $H(I_1, I_2)$ as given by the truncated non-resonant Birkhoff normal form near 
an equilibrium, so that there is exactly one torus in the pre-image, and the only singularities of 
the action map are on the boundary where one action vanishes and the corresponding torus turns 
into an isolated periodic orbit. When both actions vanish the corresponding orbit in phase space 
is the elliptic equilibrium point.
Accordingly $W(0,0)$ gives the rotation number of the equilibrium point, 
$W(0, I_2)$ and $W(I_1, 0)$ give the rotation numbers of the two families of periodic orbits 
emerging from the equilibrium point (if the Lyapunov centre theorem applies), 
and $W(I_1, I_2)$ gives the rotation number of the corresponding torus.

Fixing the energy $H(I_1, I_2) = E$ defines a line in action space, which is the projection of the 
3-dimensional energy surface onto action space. We will call this line the energy surface.
Vanishing twist occurs when the rotation number $W(I_1, I_2)$ restricted to fixed energy has 
a critical point. This can be thought of as the problem of extremizing $W$ under the constraint $H=E$.
Using a Lagrange multiplier gives the classical result that critical points occur when the surfaces $H=E$ 
and $W = const$ are tangent to each other. An analytic formula for the vanishing of twist involving 
first and second derivatives of the Hamiltonian can be obtained by implicit differentiation.
The resulting condition is
\begin{equation} 
   C(I_1, I_2) =  H_{11} H_2^2 + H_{22} H_1^2 - 2 H_{12} H_1 H_2
\end{equation}
where the subscripts on $H$ denote derivatives with respect to $I_i$.
The geometry can be visualised by considering $(H, W)$ as coordinates on action space, 
and to consider where this coordinate system has a singularity. This exactly occurs where the twist 
vanishes. We are going to use this type of picture to better understand where the twistless bifurcations
occur; namely where the line along which the twist vanishes intersects a low order rational rotation number.
In the same way we define the rotation number of the equilibrium, the periodic orbit families, and 
a general torus by evaluating $W$ at $(I_1, I_2) = (0,0)$, $I_2=0$ or $I_1=0$, or non-zero $(I_1 ,I_2)$, respectively,
we can talk about vanishing twist at the equiilibrium, the periodic orbit families, and the general torus 
by evaluating $C(I_1, I_2)$ at the corresponding actions.

In Figure~\ref{fig:rotationcontoursALL} the red curves shows where $C(I_1, I_2) = 0$, 
i.e.~where a tangency of the families $H(I_1, I_2) = E$ and $W(I_1, I_2) = const$ occurs.
Colour/shading in the figure is done according to the values of $W$, and contours of low
order rational $W$ are also shown. The gaps between these contour lines are uneven 
because the contours are selected through a Farey tree generated from $1/4$ and $1/3$, 
see, e.g.~\cite{Meiss92}.
Finally the family of thin contours lines are energy surfaces with equidistant contour lines.

The condition $C(I_1, I_2) = 0$ defines a family of twistless tori where locally the energy is a family parameter.
A family of twistless tori exists only in an integrable system. 
With a generic, non-integrable system or non-integrable perturbation of an integrable system
all (twistless) tori with rational rotation number break. Nevertheless, we will speak of the family of 
twistless tori in the CR3BP, because for most rationals the corresponding twistless bifurcations are
too small to be readily observed.

When a Poincar\'e section is done for the integrable system the energy $E$ is fixed and a 1-parameter  
family of invariant curves is obtained in the section, locally parametrised by either of the actions.
To obtain the rotation function $W(I)$ for this 1-parameter family of invariant curves in the section 
one has to eliminate one action in $W(I_1, I_2)$ using $H(I_1, I_2) = E$. 
This then gives $W(I)$ at constant energy. The natural action to use as a family parameter 
is the one that vanishes at the fixed point of the map.

%

It should be noted that the beautiful twistless bifurcation sequence as e.g.~shown in Figure~\ref{fig:twochains}
is not present in the integrable approximation given by the Birkhoff normal form.
There we merely see the collision of two tori with equal rational rotation number. 
But a small perturbation of this event brings out the full twistless bifurcation sequence.

\section{Birkhoff Normal Form}

\begin{figure*}
\centering
\includegraphics[width=0.65\textwidth]{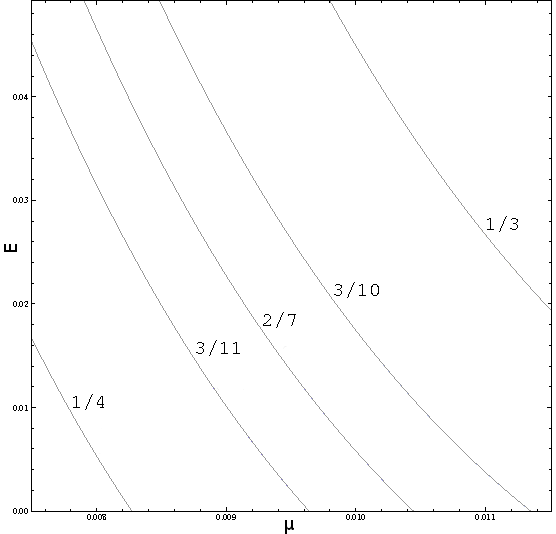}
\caption{The analytical approximations to the rotation number of the short period orbit which is the fixed point of the Poincar\'{e} map. On the horizontal axis is the mass parameter \(\mu\) between \(0.0075\) to \(0.0115\), and on the vertical axis is \( E\). The approximation is best near \(E=0\). Contours are shown for low order rationals. This compares well with the numerical results in Figure \ref{fig:collisions}.}
\label{fig:analyticalrotation}
\end{figure*}

%

\begin{figure*}
\centering
\includegraphics[ width = 5.5cm ]{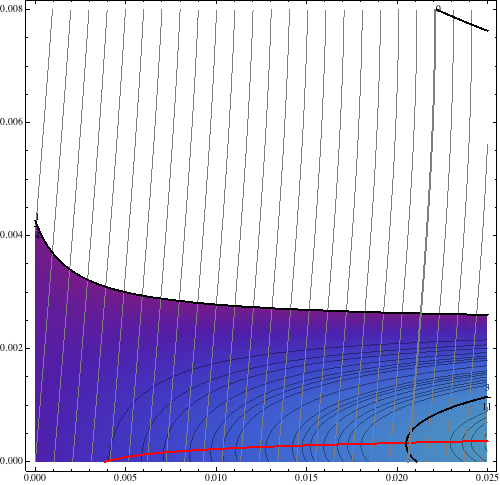} \hspace*{1ex}
\includegraphics[ width = 5.5cm ]{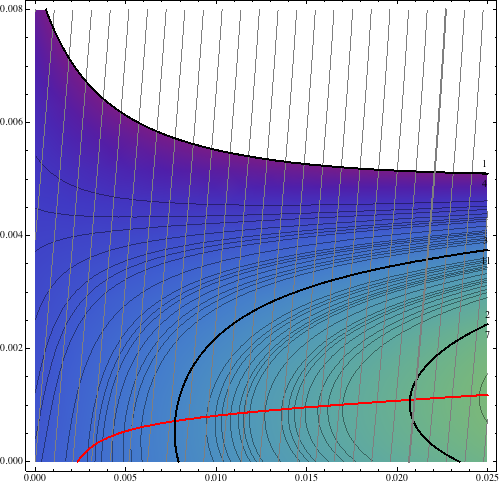}
\includegraphics[ width = 5.5cm ]{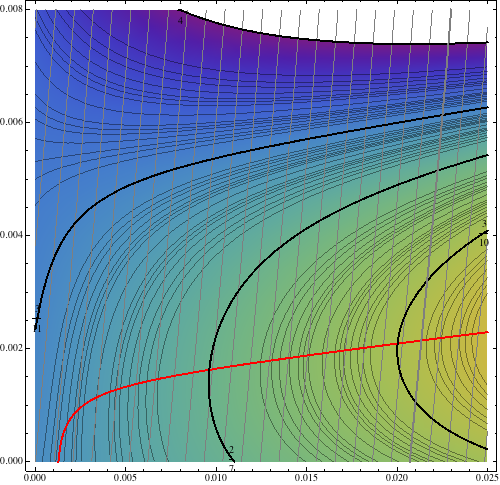} \hspace*{1ex}
\includegraphics[ width = 5.5cm ]{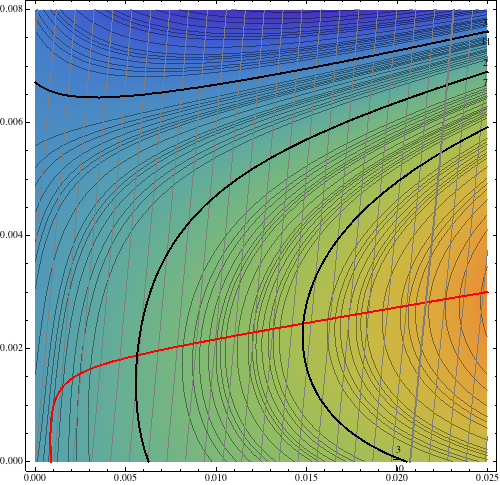}
\includegraphics[ width = 5.5cm ]{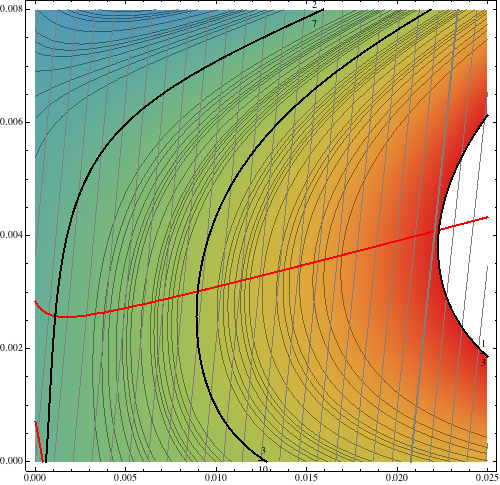} \hspace*{1ex}
\includegraphics[ width = 5.5cm ]{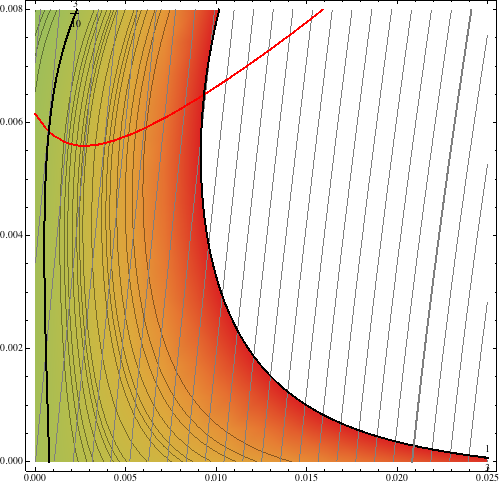}
\caption{Action vs Action plot for \(\mu=0.0086, 0.009165, 0.009723, 0.01, 0.0104, 0.01128\). 
The thick black curves indicate constant rotation number \({1}/{4}\), ${3}/{11}$, \({2}/{7}\), \({3}/{10}\) and ${1}/{3}$ with 
colour gradient from blue ($1/4$) to red ($1/3$);
thin curves of similar shape are nearby low order rational numbers. 
Nearly straight nearly vertical grey lines indicate constant energy with spacing $\Delta E = 0.01$ and $E = 0$ at the origin,
the thick grey line is $E = 0.02$.
The red curve indicates vanishing twist, visible as a tangency of these families of curves.
}
\label{fig:rotationcontoursALL}
\end{figure*}

We now consider the Birkhoff normal form of the Hamiltonian near $\mathcal{L}_4$. 
%
 For \(\mu<\mu_1\), \(\mu\neq \mu_2, \mu_3, \mu_4\)  the Birkhoff normal form is
\begin{equation} 
	\mathscr{H}(I_s,I_l)= \omega_s I_s-\omega_l I_l +\frac{1}{2}(AI^2_s+2BI_sI_l+CI_l^2)+\cdots
\end{equation}
where $\omega_s$, $\omega_l$, \(A\), \(B\), \(C\) are functions of $\mu$, 
and \(I_s\), \(I_l\), \(\phi_s\), \(\phi_l\) are the action-angle variables where the normal form is independent of the 
angle variables up to the chosen order.
The functions $A$, $B$, and $C$ were calculated explicitly by Deprit and Deprit-Bartholom\'{e} \cite{deprit67}.
 The Normal form was calculated (using computers) to sixth order in \cite{meyer86}. 
We computed the normal form to eight order but with numerical coefficients only.
Comparison with the numerical results shows that the 8th order expansion is necessary in order to get 
reasonable agreement for energy $E = 0.02$ for $\mu \in (\mu_4, \mu_3)$. Because of resonance the 
agreement deteriorates when approaching the boundaries of this interval in $\mu$.
\footnote{Notice that $\omega_s$ is not the frequency of the short period family orbit, but the limiting frequency 
of this family when it approaches the equilibrium point; similarly for $\omega_l$ and the long period family.}

As a first step we compute an analytical approximation for the rotation number of the short period family
using the analytical results of \cite{deprit67}.


The rotation number of a torus with actions $I_s$, $I_l$ is given by
\begin{equation}
	W(I_s, I_l) = -\frac{\partial H/\partial I_l }{\partial H/\partial I_s} \,.
\end{equation}
Clearly $W(0,0) = 1/r$. Keeping quadratic terms in the normal form Hamiltonian we obtain
\[
	W(I_s, I_l) =  \frac{\omega_l - B I_s - C I_l}{ \omega_s + A I_s + B I_l } \,.
\]
The short period family is obtained for $I_l = 0$ so that $E \ge 0$.
Hence
\[
       W(I_s, 0) = \frac{\omega_l - B I_s }{ \omega_s + A I_s  } , \quad 
       H(I_s, 0) = \omega_s I_s + \frac12 A I_s^2 = E \,.
\]
Eliminating $I_s$ using the second equations gives the rotation number of the short period family as a function 
of energy $E$ and parameter $\mu$, which is implicit in $\omega_s, \omega_l, A, B$.
A contour plot of the rotation number of the short period family so obtained from the analytical 4th order normal form 
is shown in Figure~\ref{fig:analyticalrotation}. 
For small $E$ the contours agree well with the numerically computed curves shown in Figure~\ref{fig:collisions}.

The reconnection bifurcations we look for as evidence of the vanishing twist occur away from the fixed point of the Poincar\'{e} map. 
To get a good approximation here we need to include higher order terms in the normal form.
We have computed the normal form with numerical coefficients up to 8th order.
The results of these computations are presented as a series of pictures in Figure~\ref{fig:rotationcontoursALL},
of the type explained in general in the previous section.

For all but the first Poincar\'e section shown in Figure~\ref{fig:psections1} there is an action-action plot 
in Figure~\ref{fig:rotationcontoursALL}. We start describing the sequence from maximal $\mu$ displayed 
last backwards towards smaller $\mu$. 

For $\mu = 0.011283$ the thick grey line $E = 0.02$ intersects the black contours of constant rotation number 
$W$ transversely, so there is no vanishing twist. Only a small range of actions have rotation number $ < 1/3$.
These correspond to the small island in the Poincar\'e section. The rotation number at the outer boundary 
of the island approaches $1/3$. The red curve indicates that there may be vanishing twist for energies
smaller than $\approx 0.06$, but the red curve does not intersect any low order rationals for $E>0$ 
so that no reconnection bifurcation will be visible. Moreover the red curve has rather large $I_s$,
and for large action the normal form may not give an accurate picture of the dynamics. 
The red curve intersects the thick black curve with $W = 3/10$ and then terminates on the long period family.
So there may be reconnection bifurcation near the long period family, but we have not been able to find them.
The reason presumably is that near the $1:3$ resonance the validity of the non-resonant Birkhoff normal 
form is rather limited.

The next action-action plot is for $\mu = 0.0104$, corresponding to the next Poincar\'e section.
The action-action plot reveals that a new segment of the red curve has been creates near the origin.
This is meaning of the value $\mu_c \approx 0.01091$, which we just passed: when the iso-energetic 
non-degeneracy condition is violated at the equilibrium $\mathcal{L}_4$ a family of twistless tori is 
created. For $\mu$ sufficiently close and smaller than  $\mu_c$ and for $|E|$ sufficiently small 
each island will have a twistless torus near the central fixed point. This is not, however, apparent 
in the Poincar\'e sections, because only when a rational rotation number is hit at sufficiently large
action does a visible reconnection bifurcation appear.
For our energy $E = 0.02$ the twistless curve is not intersecting any low order rational rotation numbers, 
it does come very close, however, to the $W=1/3$ curve. 
In the Poincar\'e section this corresponds to the pair of stable and unstable period 3 orbits 
at the outer edge of the island. Decreasing $\mu$ a little bit these period 3 orbits will annihilate in an 
saddle-centre bifurcation in which the twistless torus is created for $E= 0.02$, as described in \cite{dullin00}.
The Birkhoff normal form is slightly off in predicting the exact parameter where this happens.
It is rather surprising that it yields useful information at all so very far from the origin.

At $\mu = 0.01$ there is a twistless torus for most positive $E$, in particular for $E=0.02$, the thick grey curve
intersecting the thick red curve. The rotation number at the centre of the island (the short period family periodic orbit)
is very close to $3/10$, so this can be thought of as the birth of the $3/10$ island chain at the origin. 
However, the island chain is not yet visible in the Poincar\'e section, and the corresponding 
Poincar\'e section is not shown. 
There is a second intersection of the energy surface $E=0.02$ with the $W=3/10$ line for large action. 
Again, this is not observed in the Poincar\'e section because it is past the last torus of the island. 
An interesting bifurcation in the twistless family occurred near the origin, such that now 
there is no vanishing twist in the long period family any more, but instead only in the short period family
(for $\mu = 0.0104$ it occurred in  both periodic orbit families).
The small family of twistless tori 
near the origin cerated when $\mu = \mu_c$ now connects to the larger family that is 
present for $\mu > \mu_c$. 
\footnote{Note that for a range of $\mu$ including $\mu = 0.01$ the
red curve of twistless tori has a tangency with a particular energy near 0. This implies 
that the rotation number in the island of stability for these special parameters 
has a critical point whose second derivative also vanishes, a critical saddle point.
In principle this would create a new kind of bifurcation in which three island chains
with the same rational rotation number separated by two twistless curves interact 
with each other. However, since this occurs in a very small island (i.e.~small energy)
and not at a very rational rotation number we have not been able to find this 
bifurcation numerically.} 

Moving on to $\mu = 0.009723$, the energy surface $E=0.02$ has not changed much, 
but the rotation numbers have generally shifted to the right. Hence now there are two intersections with $W=3/10$,
with the twistless torus in between. In the Poincar\'e section this shows up as two stable period 10 
orbits with corresponding islands surrounding them and two unstable period 10 orbit with separatrices
and meandering curves between them. This is close to the reconnection bifurcation where there is 
a heteroclinic orbit connecting the unstable periodic orbits.

For $\mu = 0.009165$ 
the rotation number contours have again moved to the right, and now we are 
near the reconnection bifurcation with rotation number $2/7$.
The actual orbit corresponding to the stable period 7 orbit with smaller action 
is illustrated in Figure~\ref{fig:zspaceorbit}.
The detailed scenario of the reconnection bifurcation in this case is resolved in Figure~\ref{fig:twochains}.
The details of the reconnection bifurcation cannot be resolved from the integrable normal form, 
because it is assumed non-resonant. Hence in the integrable approximation we only see the collision 
of two rational tori with each other in a twistless torus. 
Decreasing $\mu$ below $0.009151$ the rotation number $2/7$ is not present in the island any more. 
There is a twistless curve, but its rotation number is slightly below $2/7$ and not at a (very) rational number,
and so the island appears to be foliated into invariant tori (albeit slightly distorted) completely.
A plot of the numerically computed rotation number 
for a slightly smaller $\mu$ is shown in Figure~\ref{fig:actionrotation}.
The maximum of the rotation number (i.e.~the twistless curve) has a rotation number slightly below $2/7$.
In the action-action plot Figure~\ref{fig:rotationcontoursALL} this would correspond to a figure where 
the thick black $W=2/7$ line is just to the right of the thick grey energy surface $E=0.02$.

Finally the action-action plot for $\mu=0.0086$ shows that we are near (but not exactly at) the 
$3/11$ reconnection bifurcation.
The action-action plot corresponding to the near $1/4$ resonance
is not shown since for the parameter $\mu = 0.007480$ the origin is already past the $1/4$ resonance
and for $E = 0.02$ the normal form is not valid due to the resonance.

The connection between the action-action plots in Figures~\ref{fig:rotationcontoursALL} and the global numerical 
results on reconnection bifurcations in Figure~\ref{fig:collisions} is as follows. 
Consider for example $\mu = 0.01$ in Figure~\ref{fig:collisions}.
We see that the values of the energy for which there is a reconnection bifurcation are $h\approx 0.05$ for rotation number $2/7$ 
and $h\approx0.135$ for rotation number $3/10$. In Figure~\ref{fig:rotationcontoursALL} for $\mu = 0.01$ 
these energy values can approximately be obtained by counting at which energy surface the tangency with $W=2/7$ respectively with $W=3/10$ occurs, as indicated by the red line. 
The nearly straight nearly vertical grey lines are lines of constant energy with spacing $\Delta E = 0.01$, and the tangency occurs at 
the energy values just quoted. The numerical results extend beyond the range where the normal form 
gives a good approximation.

\section{Conclusion}

We have shown that in the circular restricted three body problem near the Lagrangian equilibrium point $\mathcal L_4$ 
prominent twistless bifurcations occur for $\mu \in (\mu_4, \mu_3)$ and $0 < E < 0.1$.
The twistless torus is created in a period 3 saddle-centre bifurcation near the $1/3$ resonance through the mechanism analysed in \cite{DI03a}.
Unlike in the H\'enon map studied in \cite{dullin00}, here the twistless torus for $E=0.02$ appears to persists all the way to the 
$1/4$ resonance, as indicated by $2/7$ and $3/11$ twistless bifurcations. 
It would be interesting to study vanishing twist in an unfolding of the quadrupling bifurcation.
It is likely that other twistless bifurcations exist in this problem; in particular, for energy values \(E<0\) connecting to 
the long period family for $\mu$ close to $\mu_3$.
In particular, this may occur for the Earth-Moon system, where there is no vanishing twist in the short period family. 
The corresponding $\mu$ is already very close to $1:3$ resonance, and so 
the validity of the normal form is questionable. But it suggests that around energy $E = -0.003$
there is a twistless torus, albeit there are no low-order rationals around.
It would be interesting to study vanishing twist in the long period 
family in more detail, in particular in view of the Earth-Moon system.

%


\bibliographystyle{elsarticle-num}      
\bibliography{vanishingtwist_arxiv}   

\end{document}